\def\imat {{\rm i}} 

\documentstyle[twocolumn,floats,epsf,ifthen,aps,prb]{revtex} 

\begin{document} 
 
\draft 

\title{ 
Selective quantum evolution of a qubit state due to 
continuous measurement}

\author{Alexander N. Korotkov}
\address{ 
Department of Physics and Astronomy, State University of New York, 
Stony Brook, NY 11794-3800 
} 
\date{\today} 
 
\maketitle 
 
\begin{abstract} 
We consider a two-level quantum system (qubit) which is continuously measured 
by a detector. The information provided by the detector is 
taken into account to describe the evolution during a 
particular realization of the measurement process. We discuss the Bayesian
formalism for such ``selective'' evolution of an individual qubit 
and apply it to several solid-state setups. In particular, we show 
how to suppress qubit decoherence using continuous measurement 
and a feedback loop. 
\end{abstract}
\pacs{}
 
\narrowtext 
 

\section{Introduction}

        Studies of two-level quantum systems have acquired recently
a new meaning related to the use of this simple quantum object as an 
elementary cell (qubit) of a quantum computer.\cite{Bennett} 
This paper addresses the measurement of a qubit state,
so it necessarily touches the long-standing and still somewhat controversial
problem of quantum measurement,\cite{Wheeler,Braginsky,Giulini} 
which is known under the name of quantum state ``collapse''. 

Having in mind a solid-state realization 
of qubit (for different proposals see, e.g.\ Refs.\ 
\cite{Loss,Kane,Averin,Makhlin,Mooij}) let us 
emphasize that a realistic detector has a noisy output signal, 
so the measurement of a qubit state should necessarily have finite 
duration in order to provide an acceptable signal-to-noise ratio. 
In this situation the ``orthodox'' collapse postulate 
\cite{Neumann,Luders,Messiah} 
cannot be applied directly, since the measurement is not instantaneous. 
The necessity of a more general formalism is obvious, for example, 
in the case when the qubit ``self-evolution'' changes 
the quantum state considerably during a measurement process. 
Even if there is no self-evolution, one can wonder 
what happens with the qubit state after a partially completed measurement 
(when the signal-to-noise ratio is still on the order of unity).  
So, we need a formalism to describe the gradual qubit evolution, 
caused by the measurement process. As will be discussed later, 
the Schr\"odinger equation alone is not sufficient for the complete 
description of this evolution, and should be complemented by a slightly
generalized collapse principle.

        Continuous quantum measurement was a subject of extensive 
theoretical analysis during last two decades, and there are two main 
approaches to this problem. One approach is based on the theory of
interaction with a dissipative environment. \cite{Leggett,Weiss} 
Taking the trace over the numerous degrees of freedom of the detector, 
it is possible to obtain a gradual evolution of the measured system 
density matrix from the pure initial state to the incoherent statistical 
mixture, thus describing the measurement process.\cite{Zurek,Joos}
Since the procedure implies an averaging over the {\it ensemble},  
the final equations of this formalism are deterministic and can  be  
derived  from  the Schr\"odinger
equation alone, without any notion of state collapse. 
The success of the theory in describing many solid-state experiments 
has supported an opinion common nowadays that the collapse principle is a 
needless part of quantum mechanics.
Because of the dominance of this approach (at least in the solid-state 
community) we will call it ``conventional''.

        The other general approach to continuous quantum measurement 
(see, e.g., Refs.\ 
\cite{Carmichael,Plenio,Mensky,Gisin,Zoller,Diosi,Belavkin,Dalibard,Gisin2,%
Wiseman1,Gagen,Hegerfeldt,Griffiths,Wiseman,Calarco,Presilla,Kor-PRB,%
Doherty,Kor-LT,Goan}) 
explicitly or implicitly uses the idea of the state collapse.
Since quantum measurement is a fundamentally 
indeterministic process so that the exact measurement result is 
typically unpredictable, the approach describes the random evolution 
of the quantum state of the measured system. The important advantage 
in comparison with the conventional approach is the absence of averaging
over the total ensemble;  hence, it is 
possible to describe the evolution of an {\it individual\/} quantum system 
during a particular realization of the measurement process. The evolution 
of the measured system obviously depends on a particular measurement 
outcome; in other words, it is selected by (conditioned on) the  
measurement result. So, this approach is usually called 
the approach of {\it selective\/} or {\it conditional\/} quantum 
evolution. There is a rather broad variety of formalisms and their 
interpretations within the approach.
\cite{Carmichael,Plenio,Mensky,Gisin,Zoller,Diosi,Belavkin,Dalibard,Gisin2,%
Wiseman1,Gagen,Hegerfeldt,Griffiths,Wiseman,Calarco,Presilla,Kor-PRB,%
Doherty,Kor-LT,Goan}
 Depending on the details of the studied measurement setup 
and applied formalism, different authors 
discuss quantum trajectories, quantum state diffusion, stochastic evolution
of the wavefunction, quantum jumps, stochastic Schr\"odinger equation,
complex Hamiltonian, method of restricted path integral, Bayesian formalism, 
etc.\ (for comparison between several different ideas see, e.g. Ref.\ 
\cite{Mensky}). The approach of selective quantum evolution is relatively 
well developed in quantum optics; in contrast, 
it was introduced into the context of solid-state mesoscopics 
only recently. \cite{Kor-PRB} 

        In the present paper we continue the development of the Bayesian
formalism \cite{Kor-PRB,Kor-LT,Goan,Kor-osc,Kor-exp} for selective quantum 
evolution of a qubit due to continuous measurement. Several issues of 
the formalism derivation and interpretation are explained in more detail 
than in previous papers. A new  way of derivation is presented for 
a special case of low-transparency quantum point contact (tunnel junction) 
as a detector. We also discuss equations (briefly 
mentioned in Ref.\ \cite{Kor-LT}) for the evolution of a qubit measured 
by single-electron transistor, which go beyond the approximation used
for a nonideal detector in Ref.\ \cite{Kor-PRB}. Special attention 
is paid to a regime outside the ``weakly responding'' limit. 
Finally, we discuss the operation of a quantum 
feedback loop which can suppress the qubit decoherence caused by interaction
with the environment.

\section{Examples of measurement setup}  

        The total Hamiltonian ${\cal H}$ of a qubit continuously 
measured by a detector,  
        \begin{equation} 
{\cal H} = {\cal H}_{QB}+{\cal H}_{DET} +{\cal H}_{INT},
        \end{equation}
consists of terms describing the qubit, the detector, and their interaction.
The qubit hamiltonian, 
        \begin{equation}
{\cal H}_{QB} = \frac{\varepsilon}{2}\, (c_1^\dagger  c_1 - c_2^\dagger c_2)
        + H \, (c_1^\dagger c_2 +c_2^\dagger c_1) , 
        \label{QB}\end{equation} 
is characterized by the energy asymmetry $\varepsilon$  between two 
levels and the mixing (tunneling) strength $H$
(we assume real $H$ without loss of generality).
The Hamiltonian (\ref{QB}) is written in the basis defined by the coupling
with the detector. We will refer to mutually orthogonal states $|1\rangle$ 
and $|2\rangle$  
as ``localized'' states in order to distinguish them from the ``diagonal'' 
basis consisting of the ground and excited states, which differ in energy 
by $\hbar \Omega =(4H^2+\varepsilon^2)^{1/2}$.

\begin{figure} 
\centerline{
\epsfxsize=2in 
\hspace{0.3cm}
\epsfbox{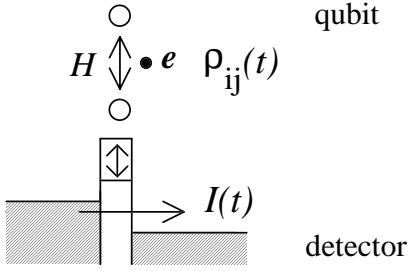}} 
\vspace{0.4cm} 
\caption{ Tunnel junction as a detector of the electron position 
in the double-dot which affects the barrier height. 
The current $I(t)$ (detector output) reflects the evolution of
the density matrix $\rho_{ij}(t)$ of the measured two-level system
(qubit). 
 }
\label{Schematic}\end{figure}

\subsection{Double-dot measured by tunnel junction} 

        Our study will be applicable to several different types of 
qubits and detectors. As the main example we consider 
a double quantum dot occupied by a single electron, the position of which 
is measured by a low-transparency tunnel junction nearby 
(see Fig.\ \ref{Schematic}). 
Basically following the model of Ref.\ \cite{Gurvitz1} 
let us assume that the tunnel barrier height depends on the location of the 
electron in either dot 1 or 2; then the current through 
the tunnel junction (which is the detector output) is sensitive 
to the electron location. In this case the detector and interaction 
Hamiltonians can be written as 
        \begin{eqnarray}
&& {\cal H}_{DET} = \sum_l E_l a_l^\dagger a_l +\sum_r E_r a_r^\dagger a_r +
\sum_{l,r} T (a_r^\dagger a_l+ a_l^\dagger a_r),  
        \nonumber\\ 
&& {\cal H}_{INT}= \sum_{l,r} \frac{\Delta T}{2} \, 
(c_1^\dagger c_1 - c_2^\dagger c_2) 
(a_r^\dagger a_l + a_l^\dagger a_r),
        \end{eqnarray}
where both $T$ and $\Delta T$ are real and their dependence on the states
in electrodes ($l,r$) is neglected. 
If the electron occupies dot 1, then the average current through
the detector is $I_1= 2\pi (T+\Delta T/2)^2 \rho_l\rho_r e^2V/\hbar$ 
($V$ is the voltage across the tunnel junction and $\rho_{l,r}$ are 
the densities of states in the electrodes) while if the measured electron 
is in the dot 2, the average current is 
$I_2=2\pi (T-\Delta T/2)^2 \rho_l\rho_r e^2V/\hbar$.

        The difference between the currents, 
        \begin{equation}
\Delta I\equiv I_1-I_2,  
        \end{equation}
determines the detector response to the electron position
(notice the different sign in the definition of $\Delta I$ used in
Ref.\ \cite{Kor-PRB}).  
        Because of the finite noise of the detector current $I(t)$, the two 
states of the system cannot be distinguished instantaneously and 
the signal-to-noise ratio gradually improves with the increase of the 
measurement duration. Let us  define  the  typical  measurement  time  
$\tau_m$ necessary 
to distinguish between two states as the time for which the signal-to-noise 
ratio is close to unity:\cite{meas_time} 
        \begin{equation}
\tau_m = \frac{(\sqrt{S_1} +\sqrt{S_2})^2}{2(\Delta I)^2}, 
        \label{tau_m}\end{equation} 
where $S_1$ and $S_2$ are the low-frequency spectral densities of the 
detector noise for states $|1\rangle$ and $|2\rangle$. 
        (As will be seen later, $\tau_m$ also determines the timescale 
for selective evolution of the qubit state due to 
measurement.) 
        For a low-transparency tunnel junction $S_{1,2}=2eI_{1,2}\coth 
(\beta eV/2)$, where $\beta$ is the inverse temperature. At sufficiently
small temperatures $\beta^{-1}\ll eV$ (we assume zero temperature 
unless specially mentioned) the detector shot noise is given  
by the Schottky formula, 
        \begin{equation}
S_{1,2}=2eI_{1,2}.
        \label{Schottky} \end{equation} 
 To avoid an explicit account of the 
detector quantum noise we will consider only processes at frequencies 
$\omega \ll eV/\hbar$ (in particular, we assume $\tau_m^{-1}\ll eV/\hbar$). 

        The major part of the paper will be devoted to the detector
in the ``weakly responding'' regime  when two states of the detector 
differ only a little (one can also call this regime ``linear'', 
while the term ``weak coupling'' is reserved for a different 
meaning), in particular, 
        \begin{eqnarray}
&& |\Delta I| \ll I_0, \,\,\,\, I_0\equiv (I_1+I_2)/2 ,
        \label{wr1}     \\
&& |S_{1}-S_2| \ll S_0, \, \,\,\, S_0\equiv (S_1+S_2)/2,  
        \end{eqnarray}  
so the typical measurement time is 
        \begin{equation}
\tau_m = 2S_0/(\Delta I)^2.
        \end{equation}
For a weakly responding detector the timescale $e/I_0$ of individual 
electron passages through the detector is much shorter than $\tau_m$,
so the current can be considered continuous on the measurement timescale. 

\subsection{Double-dot and quantum point contact}

        Besides the low-transparency tunnel junction as a detector, 
we can also consider a quantum point contact with arbitrary transparency 
${\cal T}$ which depends on the electron position in the double dot. 
This setup in the context of continuous quantum measurement 
has been extensively studied both experimentally \cite{Buks,Sprinzak} 
and theoretically. 
\cite{Aleiner,Levinson,Stodolsky,Hackenbroich,Kor-Av,Averin2} 
In spite of a somewhat different mathematical description (we will not 
write the Hamiltonian explicitly) this case is very close to the case above, 
which we prefer because of its simplicity. 
The obvious new feature is the different formula for the shot noise,
\cite{Lesovik} 
        \begin{equation} 
S_{1,2}=2eI_{1,2} (1-{\cal T}_{1,2}), 
        \label{QPC}\end{equation} 
where $I_{1,2}={\cal T}_{1,2}e^3V/
\pi\hbar$. Notice that for the quantum point contact as a detector we 
make the condition (\ref{wr1}) for weakly responding regime  
a little stronger, $|\Delta I|\ll (1-{\cal T}_{1,2})I_{1,2}$, 
so that both transmitted and reflected currents can be considered 
continuous on the measurement timescale. 

\begin{figure} 
\centerline{
\epsfxsize=2in 
\hspace{-0.1cm} 
\epsfbox{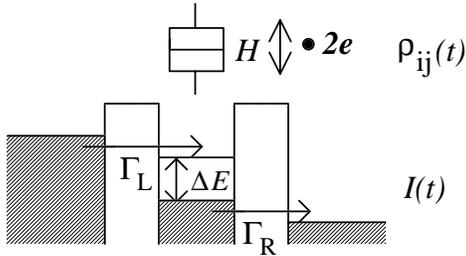}} 
\vspace{0.2cm} 
\caption{ Single-electron transistor (detector) measuring the charge state
of the single-Cooper-pair box (qubit). 
 }
\label{SET}\end{figure}

\subsection{Cooper-pair box and single-electron transistor}

        Another interesting measurement setup (Fig.\ \ref{SET}) 
introduced in Ref.\ 
\cite{Shnirman} in the context of a solid-state quantum computer,
is a single-Cooper-pair box measured by a single-electron transistor 
(a somewhat similar setup has been recently used for the experimental 
demonstration \cite{Nakamura} of quantum oscillations in the time domain). 
The qubit in this case is represented by two charge states 
of a small-capacitance 
Josephson junction. The Josephson coupling provides the matrix element 
$H$ in Eq.\ (\ref{QB}) which is assumed to be much smaller than the 
single-electron charging energy, so that only two charge states (adjusted
by the gate voltage to be close in energy) are important. The capacitively
coupled single-electron transistor is sensitive to the charge state of the 
Cooper-pair box and serves as the detector; the current $I(t)$ through the 
transistor is the measurement output.

        One can find the detailed discussion of the Hamiltonian for this 
measurement setup in Ref.\ \cite{Shnirman}. 
The qubit state affects the energy of the middle island of the 
single-electron transistor (Fig.\ \ref{SET}), so the interaction is
of ``density-density'' type: 
        \begin{equation}
H_{INT}=\frac{\Delta E}{2} \, (c_2^\dagger  c_2 - c_1^\dagger c_1) 
 \left( \sum_m a_m^ \dagger a_m -\mbox{const} \right) , 
        \label{HintSET}\end{equation}
where the factor in large brackets is the number of extra electrons on 
the transistor island. In the ``orthodox'' regime of sequential 
single-electron tunneling \cite{Av-Likh,orth} in the transistor, the 
energy change $\Delta E$ affects the rates of tunneling through 
the two tunnel junctions
and thus affects the average current $I$. In the simplest case when 
the electrons can tunnel only in a strict alternating sequence with
rates $\Gamma_L$ and $\Gamma_R$, the average currents $I_1$ and $I_2$ 
can be calculated as \cite{Av-Likh} 
        \begin{equation}
I_i=e\Gamma_{L,i}\Gamma_{R,i}/(\Gamma_{L,i}+\Gamma_{R,i}),
        \label{SET-cur}\end{equation} 
where $i=1,2$ corresponds to the charge state of the qubit.  
Well outside the Coulomb blockade range the difference between the rates is 
$\Gamma_{L,2}-\Gamma_{L,1}=-\Delta E/e^2R_L$ and 
$\Gamma_{R,2}-\Gamma_{R,1}=\Delta E/e^2R_R$, 
where $R_{L(R)}\gg \hbar/e^2$ are the resistances of tunnel junctions.

 The measurement time to distinguish between states $|1\rangle$ 
and $|2\rangle$ for this setup is given by Eq.\ 
(\ref{tau_m}), in which the spectral density of the single-electron 
transistor current can be calculated using equations of Refs.\ 
\cite{Kor-92,Kor-noise} (the Schottky formula used for this purpose 
in Ref.\ \cite{Shnirman} is valid only in a limiting case). In the special
case corresponding to Eq.\ (\ref{SET-cur}) the shot noise is given 
by the formula\cite{Kor-noise} 
        \begin{equation}
S_i = 2eI_i \, (\Gamma_{L,i}^2+\Gamma_{R,i}^2)/(\Gamma_{L,i}+\Gamma_{R,i})^2. 
        \label{SET-SI}\end{equation}

\subsection{Two SQUIDs} 

        One more solid-state realization of continuous quantum 
measurement of a qubit can be done using two flux states of a  SQUID 
as a qubit 
and another inductively coupled SQUID as a detector.\cite{Friedman}  
The corresponding Hamiltonian and calculations of the SQUID noise 
can be found, e.g.\ in Ref.\ \cite{Likh-book}. 
A minor difference in the formalism is related to the fact that 
the typical output signal from a SQUID is voltage instead of current
in the examples above.

        \section{Results of the conventional approach}

        The goal of the present paper is the analysis of a selective 
evolution of the qubit state due to continuous measurement, taking
into account the detector output $I(t)$. However, before that let us review 
the results of the conventional approach 
\cite{Gurvitz1,Aleiner,Levinson,Stodolsky,Hackenbroich,Kor-Av,Averin2,%
Shnirman} 
to this problem which does not take into account the detector output. 

        We describe the quantum state of a qubit by 
the density matrix $\rho_{ij}$ in the basis of localized states
$|1\rangle$ and $|2\rangle$, so that $\rho_{ii}$ ($\rho_{11}+\rho_{22}=1$) 
is the probability to find the system in the state ``$i$'' {\it if\/} 
an instantaneous measurement in this basis is performed, while
$\rho_{12}$ ($\rho_{21}=\rho_{12}^*$) characterizes the coherence;  
in particular, $|\rho_{12}|^2=\rho_{11}\rho_{22}$ 
corresponds to a pure state. In the conventional approach \cite{Leggett} 
the evolution  of $\rho_{ij}$ is calculated using the Schr\"odinger 
equation for the combined system including the detector and then tracing
out the detector degrees of freedom that leads to the following equations:
\cite{Gurvitz1,Aleiner,Levinson,Stodolsky,Hackenbroich,Kor-Av,Averin2,%
Shnirman} 
        \begin{eqnarray}
&&\dot{\rho}_{11}=  -\dot{\rho}_{22}=  -2\, \frac{H}{\hbar} \,
        \mbox{Im}\, \rho_{12} , 
        \label{conv1}\\
&& {\dot\rho}_{12}=  \imat \, \frac{\varepsilon}{\hbar} \,\rho_{12}+ 
        \imat \,\frac{H}{\hbar } \,
(\rho_{11}-\rho_{22}) -\Gamma_d \, \rho_{12},  
        \label{conv2}\end{eqnarray}
where the effect of continuous measurement is described by the ensemble 
decoherence rate $\Gamma_d$. (Such equations in similar problems when 
the environment causes dephasing are known for many years -- see, e.g. 
Refs.\ \cite{Leggett,Scully,Harris}.) 

        For a double-dot measured by a tunnel junction (Fig.\ 
\ref{Schematic}) the decoherence rate has been obtained in Ref.\ 
\cite{Gurvitz1}: 
        \begin{equation}
\Gamma_d = \frac{(\sqrt{I_1}-\sqrt{I_2})^2}{2e}. 
        \label{GdGur}\end{equation}
Comparing this equation with Eqs.\ (\ref{tau_m}) and (\ref{Schottky}) 
one can easily notice that $\Gamma_d$ has a direct relation to the
typical measurement time $\tau_m$: 
        \begin{equation} 
\Gamma_d =\left( 2\tau_m\right)^{-1} . 
        \label{GdTm}\end{equation} 
This relation obviously remains valid in the weakly responding regime 
when the decoherence rate can be expressed as 
        \begin{equation}
\Gamma_d = (\Delta I)^2/ 4S_0. 
        \label{GdS0}\end{equation}

        In the case of a finite-transparency quantum point contact 
as a detector 
\cite{Buks,Sprinzak,Aleiner,Levinson,Stodolsky,Hackenbroich,Kor-Av,Averin2} 
the ensemble decoherence rate has been mainly studied in the weakly 
responding regime. The most important for us result 
\cite{Buks,Sprinzak,Aleiner,Hackenbroich,Kor-Av,Averin2} is that for  
symmetric coupling Eq.\ (\ref{GdS0}) is still valid, just 
the shot noise is now given by Eq.\ (\ref{QPC}) instead of Eq.\ 
(\ref{Schottky}) (as mentioned, the temperature is zero). 
In the asymmetric case, if the phase of transmitted and reflected 
electrons in the detector is sensitive to states $|1\rangle$ and 
$|2\rangle$, then there is an extra
term in the equation for the decoherence rate, so the decoherence is faster 
\cite{Sprinzak,Stodolsky,Kor-Av,Averin2} than given by Eq.\ (\ref{GdS0}). 

        The inequality $\Gamma_d > (2\tau_m)^{-1}$ has been also obtained 
in Ref.\ \cite{Shnirman} for a single-electron transistor measuring 
a single-Cooper-pair box. The interaction Hamiltonian (\ref{HintSET}) 
allows us to relate the dephasing rate, 
$\Gamma_d =(\Delta E)^2 S_m/4\hbar^2$, 
to the low-frequency spectral density $S_m$ of the fluctuating 
number $m$ of extra electrons on the transistor central island. 
These fluctuations have been calculated in Refs.\ \cite{Kor-92,Kor-noise} 
within the framework of the orthodox theory.\cite{Av-Likh} 
In particular, assuming the weakly responding regime and the 
two-charge-state dynamics corresponding to Eqs.\ 
(\ref{SET-cur})--(\ref{SET-SI}) we obtain 
\cite{Kor-osc,Makhlin3} 
        \begin{equation}
\Gamma_d =\frac{(\Delta E)^2 \Gamma_L\Gamma_R}{\hbar^2(\Gamma_L+\Gamma_R)^3} 
        \end{equation}
(notice a different expression in Ref.\ \cite{Shnirman}). In this
case 
        \begin{equation}
2\Gamma_d\tau_m = \frac{8\Gamma_L^2\Gamma_R^2 (\Gamma_L^2+\Gamma_R^2)}
{(\Gamma_L+\Gamma_R)^2 ( \Gamma_L^2\hbar /e^2R_R-
\Gamma_R^2\hbar /e^2R_L)^2 },  
        \label{GdTmSET}\end{equation} 
so for $\Gamma_L \sim \Gamma_R$ this product is necessarily large,
$\Gamma_d \gg (2\tau_m)^{-1}$,  since $R_{R,L}\gg \hbar/e^2$. 
However, for very different tunnel rates the dephasing  can be comparable 
to $(2\tau_m)^{-1}$. Assuming $\Gamma_R \gg \Gamma_L$ one can simplify Eq.\
(\ref{GdTmSET}) to 
        \begin{equation}
2\Gamma_d\tau_m = 8 (\Gamma_L/\Gamma_R)^2 (R_Le^2/\hbar)^2. 
        \end{equation}
Formally this expression becomes less than one if $ e^2\Gamma_L R_L <
\hbar \Gamma_R /\sqrt{8}$; however, in this case the significant cotunneling 
makes the orthodox approach invalid and the quantum noise contribution 
becomes important.\cite{Kor-92} In the cotunneling 
regime (well below the Coulomb blockade threshold) $\Gamma_d$ should be 
obviously comparable to $(2\tau_m)^{-1}$ because in this case 
essentially the barrier height (the energy of the virtual 
state) is sensitive to a measured state, so the detecting principle 
becomes similar to the case of Fig.\ \ref{Schematic}. 
The inequality $\Gamma_d \geq (2\tau_m)^{-1}$ should remain valid in the 
cotunneling regime as well; this fact will be obvious from the Bayesian 
formalism. 

        The quantum backaction of a SQUID in the linear-response 
approximation was calculated in 
Ref.\ \cite{Danilov}. It was shown that the total energy sensitivity of 
a SQUID $(\epsilon_I\epsilon_V -\epsilon_{IV}^2)^{1/2}$ 
(which takes the backaction into account) is limited by $\hbar/2$. Here 
$\epsilon_V$ is the ``output'' energy sensitivity [the output signal 
of a SQUID is $V(t)$], $\epsilon_I$ describes the intensity 
of backaction noise, and $\epsilon_{IV}$ characterizes their correlation. 
From the inequality $\epsilon_I\epsilon_V \geq \hbar^2/4$ we easily get
an inequality for spectral densities: $s_Is_V \geq \hbar^2 (dV/d\Phi )^2$,
where $dV/d\Phi$ describes the SQUID response to the flux $\Phi$. 
        For the two-SQUID measurement setup considered in the present paper 
the qubit dephasing due to backaction noise is 
$\Gamma_d=(\Delta \Phi)^2s_I/4\hbar^2$ where $\Delta \Phi$ is the measured 
flux difference between two qubit states. Using the inequality above for 
the product $s_Is_V$ we obtain a lower bound for the ensemble decoherence 
rate:\cite{Averin2} 
$\Gamma_d \geq (\Delta V)^2/4s_V = (2\tau_m)^{-1}$ similar to all other 
setups discussed above. This lower bound can be achieved 
only when the SQUID sensitivity is quantum-limited. 

\vspace{0.4cm}

        Notice that the main equations (\ref{conv1})--(\ref{conv2}) 
of the conventional formalism do not depend on the detector output
$I(t)$, and so they cannot be used for the prediction of the detector current 
behavior [for generality we again choose the current as a detector output 
signal even though for a SQUID it should be changed to $V(t)$]. 
An important step towards this goal has been made in Ref.\ 
\cite{Gurvitz1} for a tunnel junction as a detector (a similar analysis
for the single-electron transistor has been performed in Refs.\ 
\cite{Shnirman,Makhlin2}). Let us divide the density matrix $\rho_{ij}$ into 
terms corresponding to different numbers $n$ of electrons 
passed through the measuring tunnel junction, $\rho_{ij}=\sum_n \rho_{ij}^n$
(only diagonal terms in $n$ are considered). Then the evolution of these
terms is given by the equations\cite{Gurvitz1} 
        \begin{eqnarray}
 \dot\rho_{11}^{\,n} = &&
        - \frac{I_1}{e} \, \rho_{11}^n + \frac{I_1}{e}\, \rho_{11}^{n-1} 
        -2\, \frac{H}{\hbar} \, \mbox{Im}\, \rho_{12}^n \, ,
        \label{Gur1}\\ 
 \dot\rho_{22}^{\,n} = && 
        - \frac{I_2}{e} \, \rho_{22}^n + \frac{I_2}{e}\, \rho_{22}^{n-1}  
        + 2\, \frac{H}{\hbar} \, \mbox{Im}\, \rho_{12}^n  \, , 
        \label{Gur2}\\ 
 {\dot\rho}_{12}^{\, n} = && 
        \imat \, \frac{\varepsilon}{\hbar} \,\rho_{12}^n+ 
        \imat \,\frac{H}{\hbar } \,(\rho_{11}^n-\rho_{22}^n) 
        -\frac{I_1+I_2}{2e} \, \rho_{12}^n    
        \nonumber \\
&&      {} + \frac{\sqrt{I_1I_2}}{e} \, \rho_{12}^{n-1} \, ,   
        \label{Gur3}\end{eqnarray}
while Eqs.\ (\ref{conv1})--(\ref{conv2}) can be derived from Eqs.\
(\ref{Gur1})--(\ref{Gur3}) after summation over $n$.

        Even though these equations couple the evolution of the system 
density matrix with the number of electrons passed through the detector,
they cannot predict the behavior of the current $I(t)$ and do not allow 
the calculation of $\rho_{ij}$ for a given realization of $I(t)$. 
        Actually, this is quite expected since the conventional
formalism describes the {\it ensemble averaged\/} evolution while the 
analysis of a particular measurement realization requires a formalism 
suitable for an {\it individual\/} quantum system. 
(The use of the conventional formalism was the reason why several recent 
attempts \cite{Stodolsky,Makhlin2,Gurvitz2} to analyze the detector current 
were not very successful.)  The analysis of a particular 
realization of the measurement process can be performed using the Bayesian 
formalism discussed in the next section.

        \section{Bayesian formalism} 

        In the Bayesian formalism (the name originates from the Bayes   
formula\cite{Bayes,Borel} for probabilities) which was derived {\it only\/}
for the weakly responding (linear) regime, the evolution of the qubit density
matrix during a particular measurement process is described by the equations 
\cite{Kor-PRB} 
        \begin{eqnarray}
\dot{\rho}_{11}= &&  -\dot{\rho}_{22}=  
-2\,\frac{H}{\hbar}\,\mbox{Im}\,\rho_{12}
         +\rho_{11}\rho_{22}\, \frac{2\Delta I}{S_0}\, [I(t)-I_0],  
        \label{Bayes1}\\ 
 {\dot\rho}_{12}= &&  \imat\, \frac{\varepsilon}{\hbar }\,\rho_{12}+ 
        \imat \, \frac{H}{\hbar } \, (\rho_{11}-\rho_{22})
        \nonumber \\ 
&& {}  -( \rho_{11}-  \rho_{22})  \frac{\Delta I}{S_0} \, 
[I(t)-I_0]\, \rho_{12} -\gamma_d \rho_{12}      
        \label{Bayes2} 
        \end{eqnarray}
(in Stratonovich interpretation -- see below),  
which replace Eqs.\ (\ref{conv1})--(\ref{conv2}) of the conventional 
formalism. Here 
        \begin{equation}
\gamma_d =\Gamma_d - \frac{(\Delta I)^2}{4S_0} \geq 0 ,
        \label{gamma_d}\end{equation} 
is the decoherence rate due to the ``pure environment'' only 
(ideal continuous measurement does not lead to this decoherence). 
One can see that $\gamma_d=0$ in the example of a tunnel junction 
as a detector, which thus can be called an ideal detector, $\eta =1$, 
where 
	\begin{equation}
\eta \equiv 1- \frac{\gamma_d}{\Gamma_d}=\frac{1}{2\Gamma_d\tau_m}.
	\end{equation}
A similar ideal situation occurs for a quantum point contact when 
$\Gamma_d=(\Delta I)^2/4S_0$, and also for the two-SQUID setup when 
the sensitivity of the measuring SQUID is quantum-limited and the output
and backaction noises are uncorrelated.
 The important prediction of the Bayesian formalism is that in such 
an ideal situation (which is experimentally 
accessible), an initially pure state of the qubit remains pure during 
the evolution; moreover, an 
initially mixed state can be gradually purified in the course of continuous 
measurement.\cite{Kor-PRB} 

        Eqs.\ (\ref{Bayes1})--(\ref{Bayes2}) allow us to calculate the 
evolution of $\rho_{ij}$ for a given measurement output $I(t)$. In order
to analyze the behavior of $I(t)$, these equations should be complemented 
by the formula 
        \begin{equation}
I(t) -I_0 = \frac{\Delta I}{2}\, (\rho_{11} -\rho_{22}) +\xi (t) ,
        \label{I(t)}\end{equation}
where $\xi (t)$ is a zero-correlated (``white'') random process with 
the same spectral density as the detector noise,\cite{rem-fr} $S_\xi =S_0$. 
The stochasticity of the detector current does not allow us to predict 
exactly the evolution of $\rho_{ij}$ in each particular realization 
of the measurement 
process; however, the formalism describes the mutual dependence of the 
stochastic evolutions of $\rho_{ij}(t)$ and $I(t)$ and thus allows us to
make experimental predictions not accessible by the conventional approach. 

        When Eq.\ (\ref{I(t)}) is substituted into Eqs.\ 
(\ref{Bayes1})--(\ref{Bayes2}), we get a system of nonlinear stochastic 
differential equations. The analysis of such equations requires 
special care, since their solution depends on the accepted 
definition of the derivative \cite{Oksendal} 
(this happens because the noise increases with the decrease of the
timescale, and so $\overline{\xi^2 dt}= \mbox{const} = S_\xi/2$ 
does not decrease with $dt$).  
 In Eqs.\ (\ref{Bayes1})--(\ref{Bayes2}) we
have used the symmetric definition, $\dot \rho (t) =\lim_{\tau\rightarrow 0}
[\rho (t+\tau /2)-\rho (t-\tau /2)]/\tau$. This is the so-called Stratonovich 
interpretation of the nonlinear stochastic equations. 
The main advantage of this interpretation is that all standard calculus 
formulas [for example, $(fg)'=f'g+fg'$] remain valid, \cite{Oksendal} 
so the intuition based on usual (nonstochastic) differential equations 
typically works well (this is the reason why we prefer the Stratonovich 
interpretation). Its other advantage is the correct limit 
in the case when the white noise term is approximated by a properly 
converging sequence of smooth functions. \cite{Oksendal} 

        However, for some purposes (e.g., for averaging over stochastic 
variables and for numerical simulations) it is more convenient to use 
another definition of the 
derivative: $\dot \rho (t) =\lim_{\tau\rightarrow 0} [\rho (t+\tau )-
\rho (t)]/\tau $. This is called the It\^o interpretation 
and it is the most commonly used interpretation in
mathematical literature on stochastic differential equations. 
 There is a simple rule of translation between the two 
interpretations:\cite{Oksendal} for an arbitrary system of 
equations 
        \begin{equation} 
\dot x_i (t) = G_i({\bf x} , t) + F_i({\bf x} ,t )\, \xi (t) 
        \end{equation} 
in Stratonovich interpretation, the corresponding It\^o equation
which has the same solution is 
        \begin{eqnarray}
\dot x_i(t) = && G_i({\bf x} , t) 
	+  \frac{S_\xi }{4} \sum_k 
        \frac{\partial F_i({\bf x} ,t)}{\partial x_k} \, F_k({\bf x} ,t)
        \nonumber \\ 
        && {} + F_i({\bf x} , t)\, \xi (t)  \, , 
        \end{eqnarray}
where $x_i(t)$ are the components of the vector ${\bf x}(t)$, $G_i$ and $F_i$ 
are arbitrary functions, and 
the constant $S_\xi $ is the spectral density of the white noise process 
$\xi (t)$. Applying this transformation to Eqs.\ 
(\ref{Bayes1}), (\ref{Bayes2}), 
and (\ref{I(t)}) we get the following equations in It\^o interpretation:
        \begin{eqnarray}
\dot{\rho}_{11} & = &  -\dot{\rho}_{22}=  -2\,\frac{H}{\hbar}\,\mbox{Im}\,
          \rho_{12}
         +\rho_{11}\rho_{22}\, \frac{2\Delta I}{S_0}\, \xi (t)\, ,  
        \label{Ito1}\\ 
 {\dot\rho}_{12} & = &  \imat\, \frac{\varepsilon}{\hbar }\,\rho_{12} +   
        \imat \, \frac{H}{\hbar } \, (\rho_{11}-\rho_{22})
        \nonumber \\ 
& - & ( \rho_{11}-  \rho_{22})  \frac{\Delta I}{S_0} \, 
\rho_{12}\, \xi (t) -\left[ \gamma_d +\frac{(\Delta I)^2}{4S_0} \right] \, 
\rho_{12} \, , 
        \label{Ito2} 
        \end{eqnarray}
while the current $I(t)$ is still given by Eq.\ (\ref{I(t)}). 
Similar equations (in a different notation) have been obtained 
in Ref.\ \cite{Gagen} for a symmetric two-level system measured by an ideal 
detector ($\varepsilon =0$, $\gamma_d=0$). Notice that the It\^o 
interpretation has been used in the majority of theories 
describing selective evolution due to quantum measurement (see 
\cite{Carmichael,Plenio,Mensky} and references therein). 

Using the It\^o interpretation it is easier to see that averaging of the 
evolution equations over the random process $\xi (t)$ (i.e.\ averaging over 
different detector outputs) leads to the conventional equations 
(\ref{conv1})--(\ref{conv2}). However, for the analysis of an individual
realization of the evolution, It\^o equations are typically less transparent
for physical interpretation. 
For example, the term $-\rho_{12}(\Delta I)^2/4S_0$ in Eq.\ (\ref{Ito2}) 
does not actually cause decoherence in an individual realization but just 
compensates the noise term proportional to $\xi^2dt$ due to the It\^o 
definition of the derivative, and so $\rho_{12}(t)$ 
does not decrease exponentially in time if $H\neq 0$. 
Similarly, the fact that the measurement tries to localize the 
density matrix in one of two states is not clear from Eqs.\
(\ref{Ito1})--(\ref{Ito2}) while it is obvious from Eqs.\ 
(\ref{Bayes1})--(\ref{I(t)}). 

        To avoid confusion due to the difference between Stratonovich 
and It\^o interpretations, it is helpful to write the exact solution of 
Eqs.\ (\ref{Bayes1})--(\ref{Bayes2}) [which is also the solution 
of Eqs.\ (\ref{Ito1})--(\ref{Ito2})] in the special case $H=0$:
        \begin{eqnarray}
&& \frac{\rho_{11}(t+\tau )}{\rho_{22}(t+\tau )} = 
\frac{\rho_{11}(t)}{\rho_{22}(t)} \,\,   
\frac{\exp [-({\overline I}(\tau )-I_1)^2\tau /S_0]}
{\exp [-({\overline I}(\tau )-I_2)^2\tau /S_0]} ,
        \label{sol1}\\
&& \frac{\rho_{12}(t +\tau )}{\left[ \rho_{11}(t+\tau )\,\rho_{22}(t+\tau )
\right]^{1/2}}=
\frac{\rho_{12}(t) \, e^{\imat \varepsilon \tau /\hbar}}
{\left[\rho_{11}(t)\,\rho_{22}(t)\right]^{1/2}}
  \,  e^{-\gamma_d \tau}, 
        \label{sol2}\end{eqnarray}
where 
        \begin{equation}
\overline{I}(\tau )\equiv \frac{1}{\tau}\int_t^{t+\tau} I(t' ) \, dt'
        \label{Iav-def}\end{equation}
is the detector current averaged over the time interval $(t, t+\tau )$.  
        These equations have clear physical meaning: Eq.\ (\ref{sol1}) 
is just the Bayes formula (see next section) while Eq.\ (\ref{sol2}) 
describes gradual decoherence due to the ``pure environment''
characterized by $\gamma_d$. 

        A useful tool for analysis of the measurement process is 
Monte-Carlo simulation of an individual process realization. For this 
purpose we can use Eqs.\ (\ref{sol1})--(\ref{sol2}) complemented by 
the simulation of evolution due to finite $H$. Let us choose a sufficiently 
small timestep $\Delta t$ (much smaller than $\hbar /H$)  
and apply the following algorithm. 
        First, for each timestep $(t, t+\Delta t)$ we pick  
the averaged current 
$\overline{I} \equiv (\Delta t)^{-1}  \int_t^{t+\Delta t} I(t') \, dt' $ 
as a random number using the probability distribution 
        \begin{eqnarray} 
P(\overline{I}) &= &  \frac{\rho_{11}(t)}{(2\pi D)^{1/2}} \, 
\exp [-\frac{(\overline{I}-I_1)^2}{2D}] 
        \nonumber \\
& +&  \frac{\rho_{22}(t)}{(2\pi D)^{1/2}} \, 
\exp [-\frac{(\overline{I}-I_2)^2}{2D}],  
        \label{Gauss1}\end{eqnarray} 
where $D=S_0/2\Delta t$. Then $\overline{I}$ is substituted into 
Eqs.\ (\ref{sol1})--(\ref{sol2}) to calculate $\rho_{ij} (t+\Delta t)$ 
from $\rho_{ij} (t)$. 
The last step of the procedure is the additional evolution during 
$\Delta t$ due to finite $H$ (rotation in the
$\rho_{11}$-$\rho_{12}$ plane).
Then the whole procedure is repeated for the next timestep $\Delta t$
and so on.  

        An alternative algorithm can be based directly on the It\^o 
equations (\ref{Ito1})--(\ref{Ito2}) which are more natural 
for numerical simulations than the Stratonovich equations because of 
the ``forward-looking'' definition of the derivative. For sufficiently 
small $\Delta t$ (now much smaller than all timescales $S_0/(\Delta I)^2$, 
$\hbar /H$, $\hbar /\varepsilon$, and $\gamma_d^{-1}$) we first calculate 
the averaged pure noise,  
$\overline{\xi} \equiv (\Delta t)^{-1}  \int_t^{t+\Delta t} \xi (t')\, dt'$, 
as a random number using the Gaussian distribution
        \begin{equation}
P(\overline{\xi }) = (2\pi D)^{-1/2} \, 
        \exp [-(\overline{\xi})^2/2D],  
        \label{Gauss2}\end{equation}
where again $D=S_0/2\Delta t$. 
Then this number is substituted into Eq.\ (\ref{Ito1}):
        \begin{eqnarray}
\rho_{11}(t+\Delta t) = && \rho_{11}(t) 
-2\Delta t (H/\hbar) \mbox{Im}\,\rho_{12}(t)
        \nonumber \\
&& {} +\rho_{11}(t)\rho_{22}(t)(2\Delta I/S_0) \overline{\xi} \, \Delta t 
        \end{eqnarray}
and similarly into Eq.\ (\ref{Ito2}). Then the updating procedure 
is repeated for the next step $\Delta t$ and so on. 
The detector current can be calculated using Eq.\ (\ref{I(t)}). 

        Both Monte-Carlo algorithms are equivalent;  
however, the first algorithm is better because it allows longer timesteps. 
The equivalence for small $\Delta t$ can be 
proven analytically using a second-order series expansion of Eqs.\ 
(\ref{sol1})--(\ref{sol2}) and has also been checked numerically.
Notice that for $\Delta t \ll S_0/(\Delta I)^2$, the current distribution  
(\ref{Gauss1}) is indistinguishable from the distribution
$P(\overline{\xi}+\Delta I(\rho_{11}-\rho_{22})/2)$ given by
Eq.\ (\ref{Gauss2}).  

\begin{figure} 
\centerline{
\epsfxsize=2.5in 
\hspace{0.3cm}
\epsfbox{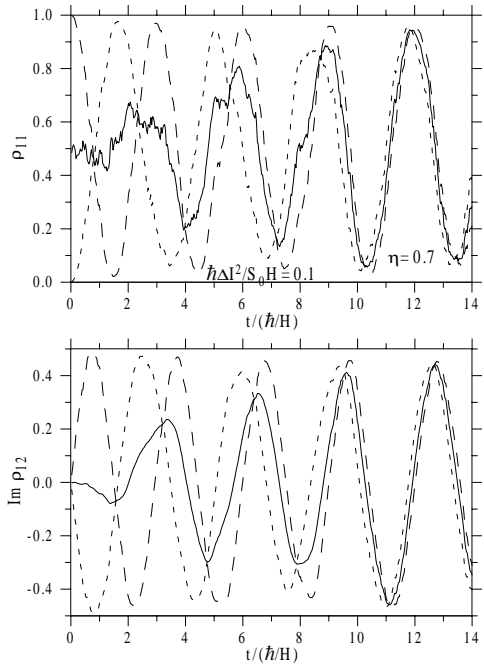}
} 
\vspace{0.4cm} 
\caption{ Solid lines: gradual purification of the qubit density matrix
$\rho (t)$ in the course of continuous measurement, starting from the 
completely incoherent state. Dashed lines show the evolution starting 
from localized states, assuming the same detector current. 
 }
\label{purif}\end{figure}

        A typical result of the Monte-Carlo simulation is shown in 
Fig.\ \ref{purif}. The solid lines show a particular realization 
of the evolution of $\rho (t)$ (diagonal and nondiagonal elements of 
the density matrix) for a symmetric qubit, $\varepsilon =0$, measured
by a detector with coupling ${\cal C}\equiv \hbar (\Delta I)^2/S_0H =0.1$
and ideality factor $\eta =0.7$. 
The real part of $\rho_{12}(t)$ is not shown since its evolution is
decoupled from $\rho_{11}(t)$ and $\mbox{Im}\,\rho_{12}(t)$. The completely
incoherent initial state is chosen, $\rho_{11}(0)=0.5$, $\rho_{12}(0)=0$.
Nevertheless, the measurement leads to the gradual onset of quantum
coherent oscillations. This happens because the measurement randomly tries 
to localize the qubit, while the finite $H$ provides oscillations when 
the state 
becomes at least partially localized. The qubit state is gradually purified, 
eventually reaching a pure state if the detector is ideal. For a nonideal 
detector (Fig.\ \ref{purif}) the state remains partially incoherent, which 
decreases the amplitude of the oscillations. 

        The qubit gradually ``forgets'' its initial state during the 
evolution and the density matrix $\rho (t)$ becomes determined mostly 
by the detector record. To illustrate this fact, the dashed lines in Fig.\
\ref{purif} show the qubit evolution calculated by Eqs.\ 
(\ref{Bayes1})--(\ref{Bayes2}) starting from two localized states and
assuming that the detector current (not shown) is exactly the same as in the 
measurement realization corresponding to the solid lines. As expected, 
after the time comparable to $\tau_m$ the dashed lines become close to 
the solid lines. 

        The tendency to qubit state localization due to measurement 
can be described quantitatively using the deterministic part of 
Eqs.\ (\ref{Bayes1}) and (\ref{I(t)}). However, because of the equation 
nonlinearity the typical localization time $\tau_{l}$ cannot 
have a unique definition. If we define it via 
an exponential-growth factor $\exp (t/\tau_l)$ for 
$\rho_{11}(t)$-evolution when $\rho_{11}$ is close to $1/2$, then
        \begin{equation}
\tau_l=2S_0/(\Delta I)^2 , 
        \end{equation}
which exactly coincides with the definition of the typical measurement time 
$\tau_m$. [If for the definition we choose the exponential-decrease 
factor $\exp (-t/\tau_l)$ when the state is almost localized, then $\tau_l$ 
would be twice smaller.]

        \section{Derivation based on Bayes formula} 

        In this section we briefly review the derivation of the Bayesian
formalism presented in Ref.\  
\cite{Kor-PRB}, which was based on the correspondence between classical 
and quantum measurements. 

        In the classical case ($H=0$, $\rho_{12}=0$) the measurement 
process can be described as an evolution of probabilities $\rho_{11}$ 
and $\rho_{22}$ which reflect our knowledge about the system 
state. Then for arbitrary $\Delta t$ (which can be comparable to $\tau_m$) 
the average current $\overline{I}$ obviously has the probability 
distribution given by Eq.\ (\ref{Gauss1}). After the measurement during 
$\Delta t$ the information about the system state has increased 
and the probabilities $\rho_{11}$ and $\rho_{22}$ should be updated using the 
measurement result $\overline{I}$ and the Bayes formula 
(\ref{sol1}), which completely describes the classical measurement. 
[The Bayes formula\cite{Bayes,Borel} says that the updated probability 
$P^*({\cal A})$ of a hypothesis ${\cal A}$ given that event ${\cal F}$ has 
happened in an experiment, is equal to $P({\cal A})P({\cal F}|{\cal A})/
\sum_{\cal B} [P({\cal B})P({\cal F}|{\cal B})]$ where $P({\cal A})$ 
is the probability before the experiment, $P({\cal F}|{\cal A})$ is the 
conditional probability of event ${\cal F}$ for hypothesis ${\cal A}$, and 
the sum is over the complete set of mutually exclusive hypotheses.] 

        The next step is an important assumption: 
in the quantum case with $H=0$ the evolution of 
$\rho_{11}$ and $\rho_{22}$ is still given by Eq.\ (\ref{sol1}) 
because there is no principal possibility to distinguish between classical
and quantum cases, performing only this kind of measurement. Even though
this assumption is quite obvious, it is not derived formally but should 
rather be regarded as a consequence of the correspondence principle.
In other words, this is the natural generalization of the collapse
postulate to the case of incomplete measurement. 
 
The comparison with classical measurement cannot describe the evolution 
of $\rho_{12}$; however, there is an upper limit: $|\rho_{12}| 
\leq [\rho_{11}\rho_{22}]^{1/2}$. Surprisingly, this inequality 
is sufficient for the exact calculation of $\rho_{12}(t)$ in the 
important special case of an ideal detector and $H=0$. 
Averaging this inequality over all possible detector 
outputs $\overline{I}$ using distribution (\ref{Gauss1}) we get the 
inequality 
        \begin{equation}
|\rho_{12}(t+\tau )| \leq [\rho_{11}(t)\rho_{22}(t)]^{1/2} \exp 
[-(\Delta I)^2\tau /4S_0]. 
        \label{inequality}\end{equation} 
On the other hand, for such averaged dynamics Eq.\ (\ref{inequality}) 
actually reaches the upper bound [see Eqs.\ (\ref{conv1})--(\ref{conv2}) and 
(\ref{GdS0})] in the cases discussed in Section III (tunnel junction, 
symmetric quantum point contact, or quantum-limited SQUID as a detector). 
This is possible {\it only\/} if in each realization of the measurement 
process
the initially pure density matrix $\rho_{ij} (t)$ stays pure all the time,
$|\rho_{12}(t)|^2=\rho_{11}(t)\rho_{22}(t)$. This fact has been the main 
point in the Bayesian formalism derivation in Ref.\ \cite{Kor-PRB}. 

        As the next step of the derivation, a mixed initial 
state has been taken into account (for $H=0$ and an ideal detector) 
using conservation of the ``degree of purity'' [Eq.\ (\ref{sol2}) 
with $\gamma_d=0$] which directly follows from a statistical consideration.
The qubit state evolution due to finite $H$ has been simply added to 
the evolution due to measurement. Finally, 
the interaction with the extra environment (which does not provide any
measurement result) has been taken into account by introducing 
the decoherence rate $\gamma_d$. 

First-order series expansion of the corresponding equations 
for $\rho_{ij}(t+\Delta t)$ leads to differential equations 
(\ref{Bayes1})--(\ref{Bayes2}). The reason why we get equations 
in Stratonovich interpretation is that the first-order expansion 
is necessarily based on the standard calculus rules which are valid 
only in this interpretation. Using a second-order expansion we can 
obtain differential equations both in Stratonovich and It\^o 
interpretations, depending on the definition of the derivative.

        \section{Alternative derivation of the formalism} 

        Let us discuss now an alternative way of deriving the Bayesian 
formalism, which is based on Eqs.\ 
(\ref{Gur1})--(\ref{Gur3}) of the conventional approach (a somewhat similar
derivation of the Bayesian formalism has been recently presented in 
Ref.\ \cite{Goan}). Since these equations have been derived \cite{Gurvitz1} 
only for the tunnel junction as a detector, we limit ourselves to this case. 

\begin{figure} 
\vspace{0.2cm}
\centerline{
\epsfxsize=2.8in 
\hspace{-0.1cm} 
\epsfbox{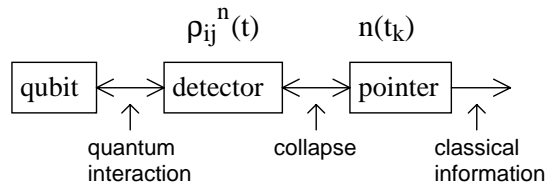} 
} 
\vspace{0.3cm} 
\caption{ 
The pointer is introduced into the model to extract the classical
signal from the detector. 
 }
\label{Pointer}\end{figure}

        Eqs.\ (\ref{Gur1})--(\ref{Gur3}) describe the coupled evolution 
of the qubit density matrix $\rho_{ij}$ and the number $n$ of electrons 
passed through the detector, considering the ``qubit plus detector'' 
as a closed system. We need to make a small but very important step 
in order to describe an individual measurement process: we need to construct 
an open system which outputs classical information to the outside. 
        For this purpose let us introduce the next stage of the 
measurement setup which will be called ``pointer'' (see Fig.\ \ref{Pointer}).
By definition, the pointer deals only with classical signals 
while quantum description is allowed for the detector. 

        Let us consider the following model. The pointer does not 
interact with the detector most of the time, however, at time 
moments $t=t_k$ ($k=1,2 ,\ldots$) the pointer measures (in simple
orthodox way) the total number $n$ of electrons passed through 
the detector. By our assumption the measured  
$n$ should be a classical number, so after each measurement by 
the pointer the number $n_k=n(t_k)$ is well defined. However, during 
the ``free'' evolution of the ``qubit plus detector'' between
the measurements by pointer the number $n(t)$ gets smeared according to 
Schr\"odinger equation, i.e.\ satisfy Eqs.\ (\ref{Gur1})--(\ref{Gur3}). 
By introducing sufficiently frequent readout (collapse) into the model 
we get the ability to describe the time dependence of the detector current.
Of course, many other 
collapse scenarios are possible, however, if we show that within some
limits the measurement process does not depend on the choice
of times $t_k$, this is a good argument justifying the generality 
of the model. 

        The collapse at $t=t_k$ can be described in the orthodox
way.\cite{Neumann,Luders,Messiah} The probability $P(n)$ to measure 
$n$ electrons passed through a detector is 
        \begin{equation}
P(n) = \rho_{11}^{n}(t_k) +\rho_{22}^{n}(t_k) .  
        \label{P(n)}\end{equation}
The measurement by pointer picks some random number $n_k$ according to 
distribution (\ref{P(n)}), however, 
after the measurement this number is already well defined and the
density matrix should be immediately updated (collapsed): 
\cite{Neumann,Luders,Messiah} 
        \begin{eqnarray}
&& \rho_{ij}^n(t_k+0) =\delta_{n,n_k} \, \rho_{ij}(t_k+0), 
        \label{collapse1}\\ 
&& \rho_{ij}(t_k+0)= \frac{\rho_{ij}^{n_k}(t_k-0)} 
{ \rho_{11}^{n_k}(t_k-0) +\rho_{22}^{n_k}(t_k-0) } , 
        \label{collapse2}\end{eqnarray} 
where $\delta_{n,n_k}$ is the Kronecker symbol.  
After that the evolution is described by Eqs.\ (\ref{Gur1})--(\ref{Gur3})
until the next collapse occurs at $t=t_{k+1}$.

        The detector current in our model has a natural averaging
during time period between $t_{k-1}$ and $t_k$ 
and can be calculated as $\overline{I}_k=e \Delta n_k/\Delta t_k$,
where  $\Delta n_k \equiv n(t_k)-n(t_{k-1})$ and 
$\Delta t_k \equiv t_k-t_{k-1}$.  Since the detector output 
is intended to reflect the evolution of the measured system, $t_k$ 
should be sufficiently frequent, in particular $\Delta t_k \ll \hbar /H$. 
For a while let us completely neglect the terms proportional to $H$ in Eqs.\ 
(\ref{Gur1})--(\ref{Gur3}) and discuss their effect later. 
Then these equations can be solved exactly. For the initial
condition $\rho_{ij}^n(0)=\delta_{n,0}\rho_{ij}(0)$ the solution is  
        \begin{eqnarray}
&& \rho_{11}^n(t)=\frac{(I_1t/e)^n}{n!}\, \exp (-I_1t/e) \, \rho_{11}(0),
        \label{P11n}\\
&& \rho_{22}^n(t)=\frac{(I_2t/e)^n}{n!}\, \exp (-I_2t/e) \, \rho_{22}(0),
        \label{P22n}\\
&& \rho_{12}^n(t)=\frac{(\sqrt{I_1I_2}t/e)^n}{n!}\, 
        \exp (-\frac{I_1+I_2}{2e}\, t +\frac{\imat \varepsilon t}{\hbar}) 
        \, \rho_{12}(0). 
        \label{P12n}\end{eqnarray} 
Similar equations describe the evolution after $k$th measurement by the 
pointer, just $t$ is shifted by $t_k$ and $n$ is 
shifted by $n_k$. Using Eqs.\ (\ref{collapse1})--(\ref{collapse2}) we 
derive the iterative equations for the qubit density matrix: 
        \begin{eqnarray}
\rho_{11} (t_{k})  = && 
\rho_{11}(t_{k-1}) \, I_1^{\Delta n_{k}} \exp (-I_1 \, \Delta t_{k}/e) 
        \nonumber \\ 
&& {} \times  [ \rho_{11}(t_{k-1}) \, I_1^{\Delta n_{k}} 
        \exp (-I_1 \, \Delta t_{k}/e) 
        \nonumber \\ 
&& {} +  \rho_{22}(t_{k-1}) \, I_2^{\Delta n_{k}} 
        \exp (-I_2 \, \Delta t_{k}/e) ]^{-1} \, , 
        \label{iter1}\\
\rho_{22} (t_{k})= && 1-\rho_{11}(t_{k}) \, , 
        \label{iter2}\\
\rho_{12}  (t_{k}) = && \rho_{12}(t_{k-1}) 
\left[ \frac{ \rho_{11} (t_{k}) \rho_{22} (t_{k}) } 
 {\rho_{11} (t_{k-1}) \rho_{22} (t_{k-1})} \right]  ^{1/2} 
        \nonumber \\ 
&&   \times \exp (\imat \varepsilon \, \Delta t_{k}/\hbar )\, , 
        \label{iter3}\end{eqnarray}
while the probability $P(n_{k})$ to get $n=n_{k}$ at $t=t_{k}$ is 
        \begin{eqnarray}
P(n_{k})  = && \frac{(\Delta t_{k}/e)^{\Delta n_{k}}}{(\Delta n_{k})!}\, 
   [ \,  I_1^{\Delta n_{k}} \exp (-I_1\Delta t_{k}/e) \, 
        \rho_{11}(t_{k-1})
        \nonumber \\
&& {} +  I_2^{\Delta n_{k}} \exp (-I_2 \Delta t_{k}/e) \, 
        \rho_{22}(t_{k-1}) ] \, . 
        \label{iter4}\end{eqnarray} 
        It is instructive to check that the averaging of $\rho_{ij}(t_{k})$  
over the result of measurement at $t=t_{k}$ gives simple equations 
        \begin{eqnarray}
 \overline{\rho_{11}(t_{k})} = && \rho_{11}(t_{k-1}), \,\,\,\,\,
\overline{\rho_{22}(t_{k})} = \rho_{22}(t_{k-1}), 
        \\
\overline{\rho_{12}(t_{k})} = && \rho_{12}(t_{k-1}) 
\exp (\imat \varepsilon \Delta t_{k} /\hbar )
        \nonumber \\
&& \times \exp [-(\sqrt{I_1}-\sqrt{I_2})^2 \Delta t_k /2e],  
        \end{eqnarray}
which are consistent with the conventional equations 
(\ref{conv1})--(\ref{GdGur}).\cite{2Gt} 

        One can easily see that Eq.\ (\ref{iter1}) can be interpreted
as the Bayes formula, while Eq.\ (\ref{iter3}) is the conservation
of the ``degree of purity'', similar to the approach reviewed above. 
The complete equivalence between Eqs.\ (\ref{iter1})-(\ref{iter4}) and 
Eqs.\ (\ref{sol1})--(\ref{Gauss1}) is achieved if 
$|\Delta I| \ll I_0$ and also the probing time 
$\Delta t_k$ is much longer than the typical time $I_0/e$ between 
individual electron passages in the detector (so that the current 
is essentially continuous). In this case the Poissonian distributions 
(\ref{P11n})--(\ref{P22n}) obviously become Gaussian, and so the probability
distributions for the current $\overline{I}=e\Delta n_k/\Delta t_k$ given by
Eq.\ (\ref{Gauss1}) and Eq.\ (\ref{iter4}) coincide. Similarly,
Eqs.\ (\ref{iter1})--(\ref{iter3}) for $\rho_{ij}$ evolution coincide 
with Eqs.\ (\ref{sol1})--(\ref{sol2}) applied to an ideal detector, 
$\gamma_d=0$. \cite{remarkNWR} 

        If the probing period is within the range $ e/I_0 \ll \Delta t_k 
\ll eI_0/(\Delta I)^2$, the evolution of $\rho_{ij}$ is smooth 
and so Eqs.\ (\ref{iter1})--(\ref{iter2}) can be written in a 
differential form which coincides with Eqs.\ (\ref{Bayes1})--(\ref{Bayes2}) 
of the Bayesian formalism with $H=0$ and $\gamma_d=0$. 
        The effect of finite $H$ can be now taken into account by the 
addition of obvious terms into Eqs.\ (\ref{Bayes1})--(\ref{Bayes2}). However,
this can be done only if $\Delta t_k \ll H/\hbar$ because in the opposite
case the terms of more than the first power in $H$ should be added into Eqs.\
(\ref{iter1})--(\ref{iter3}) indicating a nontrivial interplay between
two effects. 

        So, we have shown that in the weakly responding case, 
$\Delta I \ll I_0$, Eqs.\ (\ref{Gur1})--(\ref{Gur3}) of the conventional 
approach complemented by a sufficiently frequent readout (collapse), 
$e/I_0 \ll \Delta t_k \ll \min [eI_0/(\Delta I)^2,\, \hbar /H]$ lead to 
the equations of the Bayesian approach. The decoherence rate $\gamma_d$ is 
zero because the model\cite{Gurvitz1} describes a tunnel junction which 
is an ideal detector.

        \section{Effect of collapse due to pointer} 

        The simple model considered in the previous section allows us 
to analyze the effect of the repeated measurements by pointer on the 
qubit dynamics
in more detail and beyond the approximations of the Bayesian approach. 
First, it is important to notice that in this model the event of collapse 
at $t=t_k$ does not disturb the qubit measurement by the detector. 
More specifically, the collapse with unknown result $n_k$ is equivalent
to the absence of the collapse. To prove this fact, Eqs.\ 
(\ref{collapse1})--(\ref{collapse2}) can be averaged with the distribution
(\ref{P(n)}) that results in unity operator. 

        The absence of disturbance by pointer is because in the model there 
are no density matrix elements which couple detector states with different
number of passed electrons. Physically, this is a consequence of the 
assumption of low detector barrier transparency and infinite number 
of electrons in the detector electrodes, so 
that the ``attempt frequency'' is much larger than any collapse frequency
(for a quantum point contact the necessary condition for this assumption 
is the large resistance, $R\gg \hbar /e^2$). 
In other words, this model is intrinsically Markovian and the detector 
is classical in a sense that the passage of individual electrons 
through detector is essentially classical (not quantum) 
random process.\cite{findof} 

        The absence of the disturbance by collapse with unknown result 
does not mean, however, that we can forget about the collapse and make
it only ``at the end of the day''. Any readout from the detector 
necessarily changes the qubit state (or in other words, informs us
about the change) and thus affects the qubit evolution. 
In the limit of sufficiently often readout, 
$\Delta t_k \ll \min (e/I_1, e/I_2, \hbar /H, \hbar /\varepsilon )$, 
the evolution equations (\ref{Gur1})--(\ref{Gur3}) and 
(\ref{P(n)})--(\ref{collapse2}) simplify because at most one electron  
can pass through the detector between readouts.
During the periods of time when no electrons are passed through the 
detector,
the evolution is essentially described by Eqs.\ (\ref{Gur1})--(\ref{Gur3}) 
with $n=0$, while the frequent collapses just restore the density matrix 
normalization, leading to the continuous qubit evolution:
        \begin{eqnarray}
&& \dot\rho_{11} = -\dot\rho_{22}= 
-2\,\frac{H}{\hbar}\,\mbox{Im}\, \rho_{12} 
-\frac{\Delta I}{e}\, \rho_{11}\rho_{22} , 
        \label{QJ1}\\
&& \dot\rho_{12} =
\frac{\imat\varepsilon}{\hbar}\, \rho_{12} 
+\frac{\imat H}{\hbar} \, (\rho_{11}-\rho_{22})
+ \frac{\Delta I}{2e} (\rho_{11}-\rho_{22}) \rho_{12} .
        \label{QJ2}\end{eqnarray}
However, at moments when one electron passes through the detector, the
qubit state changes abruptly; this change is given by Eqs.\  
(\ref{iter1})--(\ref{iter3}) with $\Delta n_k=1$ and 
$\Delta t_k \rightarrow 0$: 
        \begin{eqnarray}
&&\rho_{11}(t+0) =\frac{I_1\rho_{11}(t-0)}
{I_1\rho_{11}(t-0)+I_2\rho_{22}(t-0)} \, , 
        \label{QJ3}\\
&&\rho_{22}(t+0)=1-\rho_{11}(t+0),
        \label{QJ4}\\ 
&&\rho_{12}(t+0)= \rho_{12}(t-0) \left[ \frac{\rho_{11}(t+0)\,\rho_{22}(t+0)}
{\rho_{11}(t-0)\,\rho_{22}(t-0)} \right] ^{1/2},  
        \label{QJ5}\end{eqnarray}
and can be obviously interpreted as the Bayesian update. Equations 
(\ref{QJ1})--(\ref{QJ5}) correspond to the framework of 
``quantum jump'' model.\cite{Plenio,Goan}

\begin{figure} 
\vspace{0.2cm}
\centerline{
\epsfxsize=2.8in 
\hspace{-0.1cm} 
\epsfbox{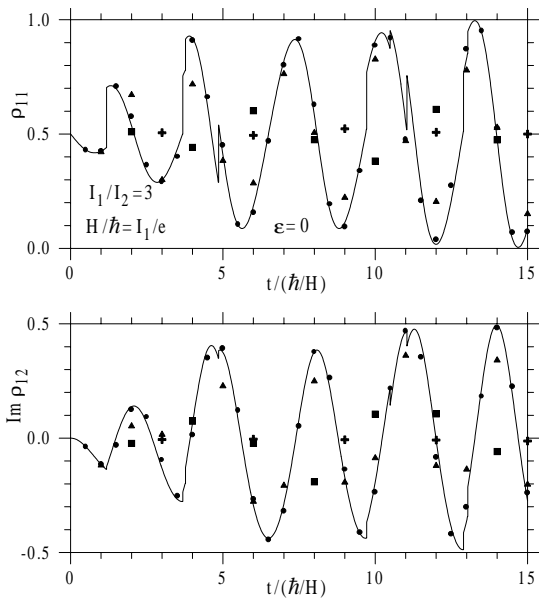} 
} 
\vspace{0.3cm} 
\caption{The lines show a gradual purification of the qubit density 
matrix $\rho (t)$ in the regime of quantum jumps (frequent detector
readout with one-electron accuracy). The dots, triangles, squares, 
and crosses correspond to finite readout periods $\Delta t_k/(\hbar /H)= 
0.5$, 1, 2, and 3 respectively. 
 }
\label{figQJ}\end{figure}

        It is easy to see that initially pure qubit state remains
pure under quantum jump evolution (\ref{QJ1})--(\ref{QJ5}) and
the density matrix is gradually purified if started from a mixed 
state. The lines in Fig.\ \ref{figQJ} show a particular realization 
of such evolution for $I_1/e=H/\hbar$, $I_1/I_2=3$, and completely incoherent 
initial state, $\rho_{11}(0)=0.5$, $\rho_{12}(0)=0$. Each discontinuity 
of curves corresponds to the passage of an electron through the detector
(the jumps of $\rho_{12}$ are typically smaller than the jumps of 
$\rho_{11}$). 
The matrix element $\rho_{11}$ always jumps up because $I_1 >I_2$ 
and so the electron passage indicates that the state $|1\rangle$ is 
somewhat more likely than state $|2\rangle$. The jumps are more 
pronounced when $\rho_{11}$
is closer to 0.5 because the jump amplitude is $\Delta \rho_{11} = 
\Delta I\rho_{11}\rho_{22}/(I_1\rho_{11}+I_2\rho_{22})$ [see Eq.\ 
(\ref{QJ3})].
The model allows us to consider 
finite ratio $I_1/I_2$ in contrast to Eqs.\ (\ref{Bayes1})--(\ref{Bayes2}) 
of the Bayesian approach. In the limit of weakly responding detector,
$|\Delta I|\ll I_0$, the amplitude of quantum jumps (\ref{QJ1})--(\ref{QJ3})
is negligible and Eqs.\ (\ref{Bayes1})--(\ref{Bayes2}) are restored 
(in this sense they describe a ``quantum diffusion'' model\cite{Goan}).
Notice, however, that equations of the Bayesian approach are applicable 
to a broader class of detectors.

\begin{figure} 
\vspace{0.2cm}
\centerline{
\epsfxsize=2.8in 
\hspace{-0.1cm} 
\epsfbox{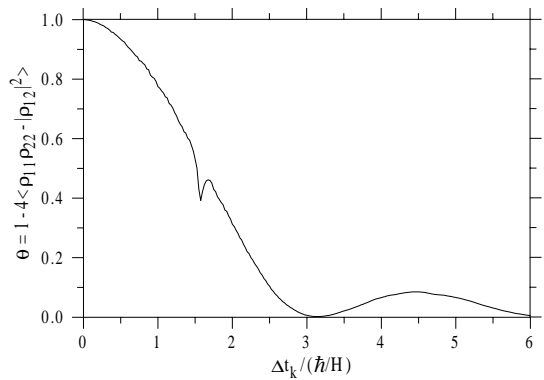} 
} 
\vspace{0.3cm} 
\caption{ 
The average qubit coherence factor $\theta$ as a function of the readout 
period $\Delta t_k$ for the measurement process shown in Fig.\ 
\protect\ref{figQJ}. 
 }
\label{figQJ2}\end{figure}

        Since the model (\ref{Gur1})--(\ref{Gur3}) describes the ideal 
detector, the qubit state in Fig.\ \ref{figQJ} eventually becomes completely 
pure. However, if the readout period $\Delta t_k$ is not sufficiently small, 
the information about the moments of electron passage through the detector
is partially lost that decreases our knowledge about the qubit state. 
In the formalism this leads to a partial decoherence of the qubit density 
matrix. The symbols in Fig.\ \ref{figQJ} (dots, triangles, squares, and 
crosses) represent the readout with several different 
periods for exactly the same realization of a measurement process 
as for the lines which represent very frequent readout. 
When the readout is still sufficiently frequent (dots), 
we can monitor the qubit evolution with a good accuracy (dots almost
coincide with the lines). However, with the increase of the readout period,   
$\rho_{11}$ becomes close to 0.5 and $\rho_{12}$ becomes close to zero, 
indicating a strongly mixed state. Figure \ref{figQJ2} shows the  
corresponding decrease of the average coherence factor $\theta \equiv 1-4  
\langle \rho_{11}\rho_{22} - |\rho_{12}|^2\rangle $ with increase of the 
readout period $\Delta t_k$ (equal time between readouts is assumed). 
The averaging is done over the readout moments
for sufficiently long realization of the measurement process. We also tried 
few other expressions which describe the density matrix coherence, all of
them show a similar dependence on $\Delta t_k$. Notice the vanishing 
coherence in Fig.\ \ref{figQJ2} when the ratio between $\Delta t_k$ and
the quantum oscillation period $\pi\hbar /H$ is close to an integer number  
(the regime of quantum nondemolition measurements\cite{Braginsky,Calarco}). 

        In the special case $H=0$ all the information about the qubit state 
is contained in the result of the last measurement by pointer. This fact can 
be easily proven by applying Eqs.\ (\ref{iter1})--(\ref{iter3}) twice 
and checking that resulting qubit density matrix does not depend on the 
result $n_1$ of the first measurement while the dependence on the second
result $n_2$ is the same as in the case of only one last measurement. 
        Similarly, the probability distribution $P(n_2)$ [see 
Eq.\ (\ref{iter4})] averaged over 
the result $n_1$ of the first measurement exactly coincides with $P(n_2)$ 
in absence of the first measurement.

        It is interesting to discuss the generalization of the model 
to the case of a low-transparency tunnel junction with finite temperature 
of electrodes. Then each of the currents $I_1$ and $I_2$ can be decomposed 
into two currents flowing in opposite directions, 
        \begin{equation}
I_i=I_i^+ -I_i^-, \,\,\, I_i^+/I_i^-= \beta eV,  
        \end{equation}
where $i=1,2$, $\beta$ is the inverse temperature, and $V$ is the voltage 
across junction. In this case Eqs.\ 
(\ref{Gur1})--(\ref{Gur3}) are replaced by the following equations:
        \begin{eqnarray}
 \dot\rho_{11}^{\,n} = && - \frac{I_1^+ +I_1^-}{e} \, \rho_{11}^n 
        + \frac{I_1^+}{e}\, \rho_{11}^{n-1} 
        + \frac{I_1^-}{e}\, \rho_{11}^{n+1} 
                \nonumber \\
&&  {}    -2\, \frac{H}{\hbar} \, \mbox{Im}\, \rho_{12}^n \, ,
        \label{Gurmod1}\\ 
 \dot\rho_{22}^{\,n} = && - \frac{I_2^+ +I_2^-}{e} \, \rho_{22}^n 
        + \frac{I_2^+}{e}\, \rho_{22}^{n-1}  
        + \frac{I_2^-}{e}\, \rho_{22}^{n+1}  
                \nonumber \\ 
&&  {}    + 2\, \frac{H}{\hbar} \, \mbox{Im}\, \rho_{12}^n  \, , 
        \label{Gurmod2}\\ 
 {\dot\rho}_{12}^{\, n} = && 
        \imat \, \frac{\varepsilon}{\hbar} \,\rho_{12}^n+ 
        \imat \,\frac{H}{\hbar } \,(\rho_{11}^n-\rho_{22}^n) 
        -\frac{I_1^+ + I_1^- +I_2^+ +I_1^-}{2e} \, \rho_{12}^n    
        \nonumber \\
&&   {}   + \frac{\sqrt{I_1^+I_2^+}}{e} \, \rho_{12}^{n-1}
        + \frac{\sqrt{I_1^-I_2^-}}{e} \, \rho_{12}^{n+1} \, .    
        \label{Gurmod3}\end{eqnarray}

        If the readout period $\Delta t_k$ is much shorter than 
$\min (e/I_i^\pm , \hbar /H)$, the detector still does not decrease 
the qubit coherence in spite of the finite temperature. However, 
if individual electron passages are not resolved, the information about 
the number of electrons passed in each direction is lost that leads 
to the qubit decoherence. In the framework of Bayesian formalism 
in the case of quasicontinuous current,  
$e/I_i^\pm  \ll \Delta t_k \ll \min [eI_i/(\Delta I_i)^2 ,\hbar /H]$, 
we can easily calculate the output current noise  
$S_0=2eI_0\coth (\beta eV/2)$ and the ensemble decoherence 
rate $\Gamma_d =\coth (\beta eV/2) (\Delta I)^2/8eI_0$ (see also the  
derivation in Ref.\ \cite{Goan}). Thus calculated detector ideality 
factor,
        \begin{equation}
 \eta =[ \tanh (\beta eV/2)]^2 , 
        \end{equation}
becomes significantly less than unity at temperatures $\beta^{-1}\agt eV$.

        \section{Detector with correlated output and backaction noises} 

        Let us assume again a weakly responding (linear) detector 
and consider the case when the output detector noise is correlated 
with the ``backaction'' noise which provides the fluctuations 
$\varepsilon (t)$ of the qubit energy level difference and thus leads
to the qubit dephasing. For example, this is the typical situation for 
a single-electron transistor as a detector.\cite{Kor-92,Kor-noise} 
In this case the knowledge of the noisy detector output $I(t)$ 
gives some information about the probable backaction noise ``trajectory'' 
$\varepsilon (t)$ which can be used to improve our knowledge of the
qubit state. The compensation for the most probable trajectory 
$\varepsilon (t)$ leads to improved Bayesian evolution equations:\cite{Kor-LT}
        \begin{eqnarray}
\dot{\rho}_{11}= &&  -\dot{\rho}_{22}=  
-2\,\frac{H}{\hbar}\,\mbox{Im}\,\rho_{12}
         +\rho_{11}\rho_{22}\, \frac{2\Delta I}{S_0}\, [I(t)-I_0],  
        \label{BayesSET1}\\ 
 {\dot\rho}_{12}= &&  \imat\, \frac{\varepsilon}{\hbar }\,\rho_{12}+ 
        \imat \, \frac{H}{\hbar } \, (\rho_{11}-\rho_{22})
        \nonumber \\ 
&& {} -( \rho_{11}-  \rho_{22})  \frac{\Delta I}{S_0} \, 
[I(t)-I_0]\, \rho_{12} 
        \nonumber \\
&& +\, \imat \, K\, [I(t)-(\rho_{11}I_1+\rho_{22}I_2)] \, \rho_{12} 
 -\tilde\gamma_d \rho_{12},     
        \label{BayesSET2} 
        \end{eqnarray}
where $K=(d\varepsilon /d\varphi )S_{I\varphi }/S_0\hbar$ characterizes 
the correlation between the noise of current $I$ through the single-electron
transistor and the noise of its central electrode potential $\varphi$
($S_{I\varphi }$ is the mutual low-frequency spectral 
density).\cite{selfcons} The term 
in square brackets after $K$ in Eq.\ (\ref{BayesSET2}) is just the 
``pure output noise'' from Eq.\ (\ref{I(t)}).  
The decoherence rate $\tilde \gamma_d $ in Eq.\ (\ref{BayesSET2}) is now 
decreased because of partial recovery of the dephasing:
        \begin{equation} 
\tilde\gamma_d =\Gamma_d - \frac{(\Delta I)^2}{4S_0} - \frac{K^2S_0}{4} \, . 
        \end{equation}
The term containing $K$ in Eq.\ (\ref{BayesSET2}) is proportional to 
the average $\varphi (t)$ for given $I(t)$. Performing ensemble averaging 
of this term [essentially, considering noise $\varphi (t)$ as uncorrelated 
with $I(t)$], we can reduce Eqs.\ (\ref{BayesSET1})--(\ref{BayesSET2}) 
to Eqs.\ (\ref{Bayes1})--(\ref{Bayes2}), while additional ensemble averaging 
over $I(t)$ leads to the conventional equations (\ref{conv1})--(\ref{conv2}).

        The obvious inequality $\tilde\gamma_d \geq 0$ (in the opposite case
the condition $|\rho_{12}|^2\leq \rho_{11}\rho_{22}$ would be violated) 
imposes a lower bound for the ensemble decoherence rate $\Gamma_d$: 
        \begin{equation}
\Gamma_d \geq \frac{(\Delta I)^2}{4S_0} + \frac{K^2S_0}{4}\, , 
        \label{Gammad}\end{equation} 
which is stronger than the inequality $2\Gamma_d\tau_m \geq 1$ 
(see Section III).

        Inequality (\ref{Gammad}) can be also interpreted in terms of
the energy sensitivity of a single-electron transistor. Let us define the 
output energy sensitivity as $\epsilon_I \equiv (dI/dq)^{-2} S_0/2C$ 
where $C$ is the total island  capacitance  and  $dI/dq  $  is  the 
response 
to the externally induced charge $q$. Similarly, let us characterize 
the backaction noise intensity by $\epsilon_\varphi \equiv CS_\varphi/2$ and
the correlation between two noises by the magnitude 
$\epsilon_{I\varphi}\equiv (dI/dq)^{-1}S_{I\varphi}/2$. Since in absence of
other decoherence sources $\Gamma_d =S_\varphi (C\Delta E/2e\hbar)^2$, 
where $\Delta E$ is the energy coupling between qubit and single-electron
transistor (see Section III), and using also the reciprocity property 
$\Delta q =C\Delta E/e=d\varepsilon /d\varphi$, we can rewrite Eq.\ 
(\ref{Gammad}) as 
        \begin{equation}
(\epsilon_I \epsilon_\varphi -\epsilon_{I\varphi}^2)^{1/2} \geq \hbar /2,  
        \end{equation}
similar to the result of Ref.\ \cite{Danilov} 
(see also Refs.\ \cite{Averin2,Zorin,Averin3,Brink}). When the 
limit $\hbar /2$ is achieved, the decoherence rate    
	\begin{equation}
\tilde\gamma_d= \frac{(\Delta I)^2}{4S_0} 
\left[\frac{\epsilon_I \epsilon_\varphi -\epsilon_{I\varphi}^2}{(\hbar /2)^2}
-1 \right]
	\end{equation}
in equations (\ref{BayesSET1})--(\ref{BayesSET2}) for the selective 
evolution of an individual qubit vanishes, $\tilde\gamma_d =0$. In this sense 
the detector is ideal, $\tilde\eta =1$, where 
	\begin{equation}
\tilde\eta \equiv 1- \frac{\tilde\gamma_d}{\Gamma_d} = 
\frac{\hbar^2 (dI/dq)^2}{S_0S_\varphi} +  
\frac{(S_{I\varphi})^2}{S_0S_\varphi} \, , 
	\end{equation} 
even though it can be a nonideal detector ($\eta <1$) by the previous 
definition, $\eta =\hbar^2 (dI/dq)^2/S_0S_\varphi $. 
[Notice a simple relation,
	\begin{equation} 
\eta =\tilde{\eta}=\frac{(\hbar /2)^2}{\epsilon_I\epsilon_\varphi} =
	\frac{1}{2\Gamma_d\tau_m} \, ,  
	\end{equation}
in absence of correlation between 
	noises of $\varphi (t)$ and $I(t)$, 
	$(S_{I\varphi})^2 \ll S_0S_\varphi$.]

A similar conclusion is also valid for other 
kinds of detectors: a quantum-limited total energy sensitivity $\hbar /2$ 
is equivalent to detector ideality, $\tilde\eta =1$.
Besides the tunnel junction, \cite{Gurvitz1} quantum point contact, 
\cite{Aleiner,Kor-Av} and SQUID,\cite{Danilov} the regime of
ideal quantum detection is also achievable by superconducting 
single-electron transisitor \cite{Zorin} and normal single-electron
transistor in cotunneling mode.\cite{Brink} 
(The resonant-tunneling single-electron transisitor\cite{Averin3} 
has ideality factor comparable, but not equal to unity.)

        \section{Quantum feedback loop}

        The Bayesian formalism allows us to monitor the evolution of
an individual qubit using weak continuous measurement, thus avoiding 
strong instantaneous perturbations. This information can be used to control
the qubit  parameters  $\varepsilon$  and  $H$  in  order  to  tune 
continuously the qubit state in such a way that the evolution follows 
the desired trajectory (a similar idea has been discussed in Refs.\ 
\cite{Wiseman,Doherty}).
This is possible even in the presence of decoherence due to the environment 
and so presents an opportunity to suppress such decoherence.

\begin{figure} 
\centerline{
\epsfxsize=2.5in 
\hspace{0.3cm}
\epsfbox{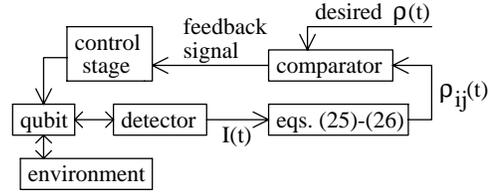} 
        }  
\vspace{0.3cm} 
\caption{ 
Schematic of continuous qubit purification using a quantum 
feedback loop. 
 }
\label{purification}\end{figure}

        Continuous qubit purification using a quantum feedback loop
\cite{Kor-exp} can be useful for a quantum computer. 
        All quantum algorithms require the supply of ``fresh'' qubits
with well-defined initial states. This supply is not a trivial problem since 
a qubit left alone for some time deteriorates due to interaction 
with the environment. The usual idea is to use the ground state which
should be eventually reached and does not deteriorate. However, 
to speed up the qubit initialization we need to increase the coupling 
with environment, which should be avoided. Another possible idea 
is to perform a projective measurement, after which the state 
becomes well-defined. However, in the realistic case the coupling
with the detector is finite, which makes projective measurement impossible. 
So, a different idea is helpful: to tune the qubit continuously in order 
to overcome the dephasing due to the environment and so keep the 
qubit ``fresh''.

        The schematic of such state purification is shown in Fig.\ 
\ref{purification}. The qubit is continuously measured by a weakly coupled 
detector, and the detector signal is plugged into 
Eqs.\ (\ref{BayesSET1})--(\ref{BayesSET2}) [or into Eqs.\ 
(\ref{Bayes1})--(\ref{Bayes2}) in a simpler case] 
to monitor the evolution of the qubit density matrix $\rho_{ij}(t)$. 
This evolution is compared with the desired evolution, and the difference is 
used to generate the feedback signal which controls the qubit 
parameters $H$ and $\varepsilon$ in order to reduce the difference with
the desired qubit state. 

        We have simulated a feedback loop designed to maintain the 
perfect quantum oscillations of a symmetric qubit ($\varepsilon =0$), 
so that the desired evolution is $\rho_{11}=[1+\cos (\Omega t)]/2$,
$\rho_{12}=\imat \sin (\Omega t)/2$ where $\Omega =2H/\hbar$. 
Let us assume an ideal detector, $\eta =1$,
so that the qubit decoherence rate $\gamma_d$ in Eqs.\ 
(\ref{Bayes1})--(\ref{Bayes2}) is due to the extra environment. The ratio 
between the decoherence rate and the ``measurement rate'' $(\Delta I)^2/4S_0$
is described by the factor $d\equiv 4S_0 \gamma_d/ (\Delta I)^2$. 

        To imitate a realistic situation the current $I(t)$ is averaged 
with a rectangular window of duration $\tau_a$ running in time, before 
it is plugged into Eqs.\ (\ref{Bayes1})--(\ref{Bayes2}). So, thus calculated 
density matrix $\rho^a(t)$ differs (a little) from the ``true'' density 
matrix $\rho (t)$ which is simultaneously simulated by the Monte-Carlo
method described in Section IV. The feedback signal is proportional
to the difference $\Delta \phi$ between the desired oscillation 
phase $\Omega (t-\tau_a/2)$ and the phase calculated as  
$\phi (t) \equiv \arctan [2\,\mbox{Im}\,\rho^a_{12}(t)/ 
(\rho^a_{11}(t)-\rho^a_{22}(t))]$. Here the time shift $\tau_a/2$ partially  
compensates the detector signal delay due to averaging. 
 The feedback signal is used to control the 
qubit tunnel barrier: $H_{fb}(t)=H[1-F\times \Delta \phi (t-\tau_d)]$ where 
$F$ is the dimensionless strength of the feedback and $\tau_d$ is an  
additional time delay ($\tau_d=0$ is preferable but not achievable in a 
realistic situation). 

\begin{figure} 
\centerline{
\epsfxsize=2.8in 
\hspace{0.0cm}
\epsfbox{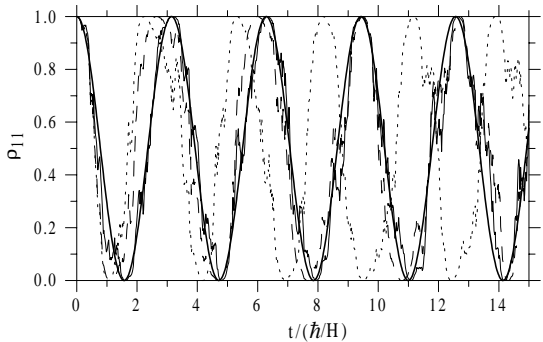} 
        }  
\vspace{0.3cm} 
\caption{Particular realizations of the qubit evolution for a quantum 
feedback loop with strength $F=3$, 0.3, and 0 (thin solid, dashed, and 
dotted lines, respectively). Thick line shows the desired evolution. 
No extra environment is present, $d=0$.
 }
\label{FB1}\end{figure}

        Figure \ref{FB1} shows typical realizations of the qubit's  
$\rho_{11}$ evolution for ${\cal C}= \hbar (\Delta I)^2/S_0H =1$, 
$\varepsilon =0$, $\tau_a=0.1\hbar /H$, $\tau_d=0.05 \hbar /H$, 
and several values of the feedback factor $F=0$, 0.3, and 3. 
No extra environment is assumed, $d=0$.  The qubit evolution starts 
from a localized state: $\rho_{11}(0)=1$, $\rho_{12}(0)=0$, and the 
desired evolution is shown by the thick solid line.  
Without a feedback ($F=0$) the phase of quantum oscillations randomly
fluctuates (diffuses) in time. However, for sufficiently large $F$ 
the feedback ``locks'' the qubit evolution and makes it close 
to the desired one. Further increase of $F$ decreases the difference 
between the actual and desired evolution. When $F$ is too strong,
the feedback loop becomes unstable.  Overall, the behavior of 
this quantum feedback loop is similar to the behavior of a traditional
classical feedback loop. In particular, we have checked that the increase 
of the averaging time $\tau_a$ and/or delay time $\tau_d$ eventually leads 
to synchronization breakdown. 
	A decrease of the detector coupling ${\cal C}$ decreases 
the evolution disturbance due to measurement and allows more accurate
tuning of quantum oscillations; on the other hand, in this case the 
feedback control becomes weaker and slower. 

\begin{figure} 
\centerline{
\epsfxsize=2.8in 
\hspace{0.0cm}
\epsfbox{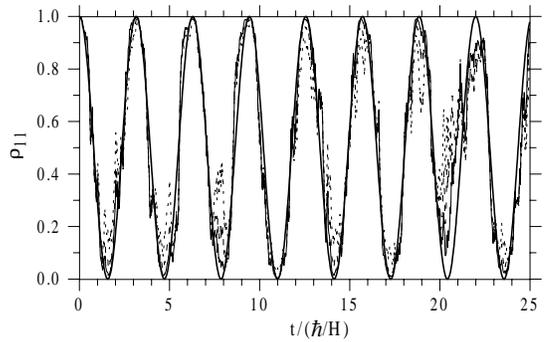} 
        }  
\vspace{0.3cm} 
\caption{Operation of the quantum feedback loop (particular realization) 
for several decoherence rates due to extra environment, 
$d= \gamma_d \times 4S_0 / (\Delta I)^2 = 0.3$, 1, and 3 (thin solid, dashed,
and dotted lines, respectively). Thick solid line is the desired
evolution. $F=3$, $\hbar (\Delta I)^2/S_0H =1$, 
$\varepsilon =0$, $\tau_a=0.1\hbar /H$, $\tau_d=0.05 \hbar /H$. 
 }
\label{FB2}\end{figure}

         Qubit decoherence due to the presence of an extra environment 
prevents complete purification of the quantum oscillations so that
the average qubit coherence factor $\theta$ becomes less than 100\%. However,
if the qubit coupling with the detector is stronger than the coupling
with its environment, 
$d\alt 1$, the feedback loop still provides qubit evolution 
quite  close  to  the  desired  one  (see  Fig.\  \ref{FB2}).   
Most noticeably,
the phase of quantum oscillations does not diffuse far from the
desired value $\Omega t$.  So, for example, the spectral density 
of these oscillations has a delta-function shape at frequency $\Omega$ 
(with exponentially small width) in contrast to the maximum value 
of 4 for the peak-to-pedestal ratio in the case of 
quantum oscillations without feedback.\cite{Kor-osc,Kor-Av}

        \section{Discussion} 

        The Bayesian formalism discussed in this paper presents
(as any formalism of selective quantum evolution
\cite{Carmichael,Plenio,Mensky,Gisin,Zoller,Diosi,Belavkin,Dalibard,Gisin2,%
Wiseman1,Gagen,Hegerfeldt,Griffiths,Wiseman,Calarco,Presilla,Kor-PRB,%
Doherty,Kor-LT,Goan})  a controversy 
in interpretation. First of all, a natural question is how it is 
possible that the qubit density matrix evolution can be described
simultaneously by the conventional equations (\ref{conv1})--(\ref{conv2})
and the Bayesian equations (\ref{Bayes1})--(\ref{Bayes2}) [and even 
also by the improved Bayesian equations (\ref{BayesSET1})--(\ref{BayesSET2})].
Which equations are correct? The answer is: all are correct depending on
the problem considered.
 
        If only the ensemble evolution is studied (for example,
the ensemble of particles is measured, as in typical nuclear magnetic 
resonance experiments) then the conventional approach is completely
sufficient. It is also possible to use the Bayesian equations; however,
they should be averaged over all possible measurement results, 
after which they coincide with the conventional equations. 
So, the selective approach does not have 
real advantages for the study of the averaged evolution (besides 
a significant computational gain in some cases
\cite{Carmichael,Plenio,Mensky}). 
There is still no advantage even for the majority of experiments with 
individual quantum systems (see examples in Section II) {\it if\/} 
the averaging is done 
over a number of repeated experiments, disregarding the results of
individual measurements (more exactly, when not more than one number 
is recorded as a result of each run). 

        The principal advantage of the selective evolution approach 
arises for continuous measurement of an individual quantum system when 
the continuous detector output $I(t)$ is recorded 
(or at least two numbers are recorded in each run). 
In this case the selective approach gives the possibility to make 
experimental predictions, unaccessible for the conventional 
approach. The proposals of such experiments with solid-state qubits 
have been discussed, 
for example, in Ref.\ \cite{Kor-PRB} for a one-detector setup and
in Ref.\ \cite{Kor-exp} for a two-detector setup (the latter experiment
seems to be realizable at the present-day level of solid-state
technology). 

        In this case the density matrices calculated by the conventional 
and Bayesian equations are significantly different. However, they do not
contradict each other but rather the Bayesian-calculated density matrix
is more accurate than the conventional counterpart. For example,
there are no situations when two approaches predict different pure states
of the qubit -- then it would be possible to prove experimentally than one 
of the approaches is wrong. Instead, in a typical situation the conventional 
equations give a significantly mixed state (so, essentially no predictions 
are possible) while the Bayesian equations give a pure state (and so some 
predictions with 100 \% certainty are possible). 
        A similar relation holds between Bayesian equations 
(\ref{Bayes1})--(\ref{Bayes2}) and the improved Bayesian equations 
(\ref{BayesSET1})--(\ref{BayesSET2}): the latter give a more accurate
description of qubit evolution and allow us to make more accurate 
predictions.  

        The difference between density matrices calculated in
different approaches can be easily understood if we treat density 
matrix not as a kind of ``objective reality'' but rather as our knowledge 
about the qubit state (in accordance with orthodox interpretation
of quantum mechanics). Then it is obvious that since Bayesian equations
take into account additional information [detector output $I(t)$], 
they provide us with a more accurate description of the qubit state than 
the conventional equations. 

        Another controversial issue is the state collapse due to 
measurement (here it is more appropriate to mention the mathematical 
formulation by  L\"uders \cite{Luders,Gisin} rather than by von Neumann 
\cite{Neumann}). The conventional equations are derived 
without any notion of collapse while the derivation of Bayesian 
equations requires either implicit or explicit (as in the model 
of Section VI) use of the collapse postulate. Philosophically, the
collapse postulate is almost trivial: when the result of the measurement 
becomes available, we know for sure that the state of the measured 
system has changed consistently with the measurement result
(even though it is generally impossible to predict the result with 
certainty). 
       In spite of being trivial, this postulate in the author's opinion 
cannot even in principle be derived dynamically by the deterministic 
Schr\"odinger equation because of the intrinsic randomness of the measurement
result. In other words, the measurement process cannot be 
described by the Schr\"odinger equation alone because this equation is 
designed for closed systems while a quantum object under measurement 
is always an open system (even including the detector), since the 
measurement information is output to the outside world. 
(The incompatibility between quantum mechanics and ``macrorealism'' has
been discussed, e.g., in Ref.\ \cite{Leggett2}.) 

        Following the orthodox (Copenhagen) interpretation, we can regard
collapse not as a real physical process but rather as a convenient
formal tool to get correct experimental predictions.
In the author's opinion this tool is still irreplaceable (if we leave aside
the many-worlds interpretations\cite{Everett,DeWitt}) for the complete 
description of the quantum realm. (Of course, in many cases the collapse 
postulate is not necessary as, for example, for the description of 
decoherence due
to interaction with the environment -- this problem has been solved with 
great success by the conventional approach.) 

        Bayesian equations predict several quite counterintuitive 
results. For example, even for a qubit with an infinite barrier between
localized states, $H=0$, the continuous measurement by an ideal detector 
leads to a gradual ``flow'' of the wavefunction between the states 
(for an initially coherent qubit). The interpretation of this effect is
rather difficult if we treat the wavefunction as objective reality; 
in contrast, there is no problem with the orthodox interpretation. 
Most importantly, experimental observation of such effects in solid-state
qubits is coming into the reach of present-day technology.
These experiments would be extremely important not only for better 
understanding of the foundations of quantum theory, but could be also 
useful in the context of quantum computing. 

        The author thanks S.\ A.\ Gurvitz, D.\ V.\ Averin, H.-S.\ Goan, 
M.\ H.\ Devoret, A. Maassen van den Brink,  G. Sch\"on, and Y.\ Nakamura 
for useful discussions.


\begin{references}
\bibitem{meas_time} Actually, for $S_1 \neq S_2$ the signal-to-noise 
        ratio corresponding to $\tau_m$ defined by Eq.\ (\protect\ref{tau_m})
        is not exactly unity; however, it has the desired asymptotic behavior 
        $(t/\tau_m)^{1/2}$ at $t/ \tau_m \gtrsim |\ln (S_1/S_2) |$. 
        Notice that a reliable measurement 
        result can be obtained only after several times $\tau_m$ ($2\tau_m$ 
        gives 98\% reliability, $3\tau_m$ gives 99.9\%, etc.).
\bibitem{orth} By coincidence, the word ``orthodox'' is used for both 
	Copenhagen interpretation of quantum mechanics and for 
	the semiclassical theory of correlated single-electron tunneling.
\bibitem{rem-fr} Notice that the noise spectral density $S_0$ is treated
	in the Bayesian formalism as frequency-independent. This implies
	that we do not consider frequencies higher than the typical
	frequency $\omega_0$ above which $S_0$ becomes frequency-dependent.
\bibitem{2Gt} In a more general case of an arbitrary detector with the
	possibility to observe each passing electron, Eqs.\ 
	(\protect\ref{iter1})--(\protect\ref{iter2}) remain valid, so
	using the inequality $|\rho_{12}|^2\leq \rho_{11}\rho_{22}$ and
	ensemble averaging we obtain the lower bound 
	$\Gamma_d \geq (\sqrt{I_1}-\sqrt{I_2})^2/2e$. If the Schottky
	formula is still applicable, $S_i=2eI_i$, this makes
	the general inequality $2\Gamma_d\tau_m \geq 1$ valid even
	outside the weakly responding regime.
\bibitem{remarkNWR} At the first glance the condition 
        $\Delta t_k \gg e/I_{1,2}$ seems to be sufficient for the 
        equivalence between the two results. 
        However, if $I_1$ and $I_2$ are significantly 
        different, then the averaging of $\rho_{12}(t)$ gives different
        results in the two models (see remark 28 in Ref.\ 
	\protect\cite{Kor-PRB}).
        This is because the tails of Poissonian and Gaussian distributions 
        are different even when the central parts coincide well.
\bibitem{findof} In a model with an effectively finite number of  
	detector degrees of freedom the collapse by the pointer would 
	disturb the interaction between qubit and detector; in particular, 
	sufficiently frequent collapse could lead to Quantum Zeno freezing 
	of the detector state.
	For example, for a finite-transparency quantum point contact the 
	``attempt frequency'' $eV/\hbar$ is always comparable to the typical
	frequency $I/e$ of electron passage. As a consequence, a readout 
	often and strong enough to observe individual electrons, 
	$\Delta t_k \alt e/I$, would necessarily disturb the measurement 
	process.
\bibitem{selfcons} The small shift of the detector operating point
	for two localized qubit states can affect the energy $\varepsilon$, 
	which therefore should be defined self-consistently.
\end{references}
\end{document}